\documentclass{article}

\PassOptionsToPackage{square, numbers, sort}{natbib}

\usepackage[final]{neurips_2019}

\usepackage[utf8]{inputenc} 
\usepackage[T1]{fontenc}    
\usepackage[hidelinks]{hyperref}
\usepackage{url}            
\usepackage{booktabs}       
\usepackage{amsfonts}       
\usepackage{amsmath}
\usepackage{nicefrac}       
\usepackage{microtype}      
\usepackage{color}
\usepackage{tikz}
\usepackage{pgfplots}
\usepackage{enumitem}

\definecolor{g-blue}{HTML}{2E86C1}
\definecolor{g-red}{HTML}{B03A2E}
\definecolor{g-purple}{HTML}{AF7AC5}

\title{Document Ranking with a Pretrained Sequence-to-Sequence Model}

\author{%
  Rodrigo Nogueira,\thanks{Equal contribution.}~~~Zhiying Jiang,$^*$ and Jimmy Lin \\[1ex]
  David R. Cheriton School of Computer Science\\
  University of Waterloo\\
}

\begin{document}

\maketitle

\begin{abstract}
This work proposes a novel adaptation of a pretrained sequence-to-sequence model to the task of document ranking.
Our approach is fundamentally different from a commonly-adopted classification-based formulation of ranking, based on encoder-only pretrained transformer architectures such as BERT.
We show how a sequence-to-sequence model can be trained to generate relevance labels as ``target words'', and how the underlying logits of these target words can be interpreted as relevance probabilities for ranking.
On the popular MS MARCO passage ranking task, experimental results show that our approach is at least on par with previous classification-based models and can surpass them with larger, more-recent models.
On the test collection from the TREC 2004 Robust Track, we demonstrate a zero-shot transfer-based approach that outperforms previous state-of-the-art models requiring in-dataset cross-validation.
Furthermore, we find that our approach significantly outperforms an encoder-only model in a data-poor regime (i.e., with few training examples).
We investigate this observation further by varying target words to probe the model's use of latent knowledge.
\end{abstract}

\section{Introduction}

A simple, straightforward formulation of ranking is to convert the task into a classification problem, and then sort the candidate items to be ranked based on the probability that each item belongs to the desired class.
Applied to the document ranking problem in information retrieval---where given a query, the system's task is to return a ranked list of documents from a large corpus that maximizes some ranking metric such as average precision or nDCG---the simplest formulation is to deploy a classifier that estimates the probability each document belongs to the ``relevant'' class, and then sort all the candidates by these estimates.

Deep transformer models pretrained with language modeling objectives, exemplified by BERT~\cite{devlin-etal-2019-bert}, have proven highly effective in a variety of classification and sequence labeling tasks in NLP.
Nogueira and Cho~\cite{Nogueira:1901.04085:2019} were the first to demonstrate its effectiveness in ranking tasks.
Since it is impractical to apply inference to {\it every} document in a corpus with respect to a query, these techniques are typically applied to {\it rerank} a list of candidates.
In a typical end-to-end system, these candidates are from the results of a keyword search based on a ``classic'' IR scoring function such as BM25~\cite{robertson1995okapi}.
This gives rise to the standard multi-stage pipeline architecture of keyword retrieval followed by reranking using one or more machine learning models~\cite{Asadi_Lin_SIGIR2013,Nogueira_etal_arXiv2019_multistageBERT}.

The contribution of this work is to adapt a pretrained sequence-to-sequence model (in our case, T5~\cite{raffel2019exploring}) to the task of document reranking.
To our knowledge, this is a novel use of this class of models that has not been previously described in the literature.
In a data-rich regime, with lots of training examples, our method can outperform a pure classification-based encoder-only approach.
However, the sequence-to-sequence model appears to be far more data-efficient:\ our approach shines in a data-poor regime and significantly outperforms BERT with limited training examples.
The main advantage of our approach, we believe, is that by ``connecting'' fine-tuned latent representations of relevance to related output ``target words'', we can exploit the model's latent knowledge (e.g., of semantics, linguistic relations, etc.)\ that has been honed through pretraining.
We describe probing experiments that attempt to verify our intuitions by deliberately altering the target words to capture different aspects of ``semantic relatedness''.

\section{Method}

Our reranking method is based on T5~\cite{raffel2019exploring}, which is a sequence-to-sequence model that uses a similar masked language modeling objective as BERT to pretrain its encoder--decoder architecture.
In this model, all target tasks are cast as sequence-to-sequence tasks.
For our task, the input sequence is:
\begin{equation}
\text{Query: } q \text{ Document: } d \text{ Relevant:}
\end{equation}
\noindent where $q$ and $d$ are the query and document texts, respectively.
The model is fine-tuned to produce the words ``true'' or ``false'' depending on whether the document is relevant or not to the query.
That is, ``true'' and ``false'' are the ``target words'' (i.e., ground truth predictions in the sequence-to-sequence transformation).

At inference time, to compute probabilities for each query--document pair (in a reranking setting), we apply a softmax only on the logits of the ``true'' and ``false'' tokens.
Hence, we rerank the documents according to the probabilities assigned to the ``true'' token.
We arrived at this particular approach after some trial and error.
Other approaches, for example, reranking documents according to the logit of the ``true'' token or using logits of all tokens to compute the softmax, were not effective, i.e., the retrieval metrics were close to zero.

Note that T5 tokenizes sequences using the SentencePiece model~\citep{kudo2018sentencepiece}, which might split a word into subwords.
We choose target words (``true'' and ``false'') that are represented as single tokens; thus each class is represented by a single logit.
In the case where target words are split in multiple subwords, we would need a method to aggregate their logits into a single score; it is best to avoid this complexity in the design of the reranking setup.

\section{Experimental Setup}

\subsection{Datasets}

We use the following datasets in our experiments:

\smallskip \noindent {\bf MS MARCO passage}~\cite{nguyen2016ms} is a passage ranking dataset with 8.8M passages obtained from the top 10 results retrieved from the Bing search engine (from 1M queries).
Note that for terminological consistency, we refer to each ``unit'' in the corpus as a document, even though they are in reality paragraph-length passages.
The training set contains approximately 500k pairs of query and relevant documents.
Each query has one relevant passage, on average.
Non-relevant documents for training are also provided as part of the training dataset.
The development and test sets contain approximately 6,900 queries each, but relevance labels are only publicly available for the development set.
We have not (yet) submitted our runs to the official MS MARCO leaderboard because our primary goal in this work is to conduct initial comparisons between T5 and BERT-based models.
As a matter of good experimental practice, we limit official submissions as to not ``probe'' the unseen test set unnecessarily.
After sufficient model refinement, we will proceed with official submissions to verify the quality of our models.

\smallskip \noindent {\bf Robust04}~\cite{voorhees2004overview} represents the test collection from the TREC 2004 Robust Track.
It comprises 250 queries, with relevance judgments on a collection of 528K documents (TREC Disks 4 and 5), whose average length is 2,800 characters or 460 words.
We use the topic ``titles'' (short keyword phrases, much like the input to a search engine) as queries to our bag-of-words retrieval methods (see Section~\ref{section:retrieval-models}) and the topic ``descriptions'' (sentence-length statements of information needs) as input to our sequence-to-sequence models.
These topic descriptions are more similar to MS MARCO's natural language questions, and others have found that using them improves the effectiveness of pretrained reranking models \cite{padakirethinking}.
We do not train our models on this dataset, and use all its queries and relevance judgments as a held-out test set; thus, our evaluation adopts a zero-shot transfer setting.

\subsection{Training and Inference}
\label{section:training-and-inference}

We fine-tune our T5 models (base, large, and 3B) with a constant learning rate of $10^{-3}$ for 100k iterations with class-balanced batches of size 128.
To simplify our training procedure (and related hyperparameters) as well as to eliminate the need for convergence checks, we simply trained for a fixed number of iterations, selected based on the computational demands of our largest model and the (self-allotted) time for running experiments.
We report results using the model state at the final checkpoint.
This procedure is consistent with the advice of Kaplan et al.~\cite{Kaplan:2001.08361:2020} and recommendations by Dodge et al.~\cite{dodge2019show}, since we quantify effectiveness for a particular computational budget.
We did not experiment with T5-11B due to its computational cost.

We use a maximum of 512 input tokens and one output token. 
In the MS MARCO passage dataset, none of the inputs have to be truncated when using this length.
We use Google's TPU v3s to train and run inference.
Training T5 base, large, and 3B take approximately 12, 48, and 160 hours overall, respectively, on a single TPU.

We use greedy decoding during inference. 
Since we only use the logits of one decoding step, beam search or top-$k$ random sampling~\cite{fan2018hierarchical} would give the same results as greedy decoding.

Because Robust04 contains full-length documents, it is not feasible to directly apply our method to the {\it entire} text at once due to the length restrictions of the model.
To address this issue, we first segment each document into passages by applying a sliding window of 10 sentences with a stride of~5.
We then obtain a relevance probability for each passage by classifying it independently.
We select the highest probability among these passages as the relevance probability of the document.

\subsection{Baselines}
\label{section:retrieval-models}

We compare our method against the following baselines:

\smallskip \noindent {\bf BM25}:\
For baseline bag-of-words retrieval, we use the BM25 implementation in the Anserini open-source IR toolkit~\cite{yang2017anserini},\footnote{\url{http://anserini.io/}} which is based on Lucene.
We adopt all the default settings.
At inference time, we retrieve the top 1000 documents per query.

\smallskip \noindent {\bf BM25+RM3}:\ 
To examine the effects of query expansion, we applied the BM25+RM3 model as described in Yang et al.~\cite{yang2019critically}, where it is shown to be a competitive baseline for (pre-BERT) neural ranking models.
We use the implementation in Anserini, with all default settings.

\smallskip \noindent {\bf BM25+BERT-large}:\ 
We additionally compare our method against the BERT-large condition from Nogueira et al.~\cite{Nogueira_etal_arXiv2019_multistageBERT}, which is a two-stage pipeline with bag-of-words retrieval (BM25) followed by a BERT reranker.
Architecturally, it is the same as our method, the only difference being BERT vs.\ T5 as the reranking model.
Nogueira et al.~\cite{Nogueira_etal_arXiv2019_multistageBERT} can be characterized as the baseline of the best methods from the official MS MARCO passage leaderboard; all higher-ranked submissions can be described as improvements upon this basic approach, and thus it represents a fair yet competitive comparison point.
Note that we did not apply reranking on top of BM25+RM3 because RM3 is known to reduce effectiveness when evaluated using these relevance judgments~\cite{Nogueira_etal_arXiv2019}.

\section{Results and Analysis}

\subsection{Main Results}

Main results on the development set of the MS MARCO passage retrieval task are shown in Table~\ref{tab:msmarco}, comparing BERT-large~\cite{Nogueira:1901.04085:2019} and T5 models of different sizes.
Results in \textbf{bold} are significantly better ($p<0.01$) than BERT-large, based on the Student's paired $t$-test.
Note that the training of T5-3B did not appear to have converged yet, even after exhausting our computational budget (see Section~\ref{section:training-and-inference}).
In other words, we suspect that T5-3B remains under-trained at the checkpoint we used for evaluation, and it is likely that effectiveness would continue to rise given more computational resources.
Larger models outperforming smaller ones is an expected trend, and with T5-11B we might observe even higher MRR@10; unfortunately, we were not able to run these experiments due to their high computational costs.

\begin{table}[t]
\begin{center}
\begin{tabular}{l|c}
& MRR@10 \\
\toprule
BM25 & .184 \\
\ \ \  + BERT-large~\cite{Nogueira_etal_arXiv2019_multistageBERT} & .372 \\
\ \ \  + T5-base & .363 \\
\ \ \  + T5-large & \textbf{.383} \\
\ \ \  + T5-3B & \textbf{.382} \\
\bottomrule
\end{tabular}
\vspace{2mm}
\caption{Results on the development set of MS MARCO passage.}
\label{tab:msmarco}
\end{center}
\end{table}

\begin{table}
\begin{center}
\begin{tabular}{l|ccc}
& AP & P@20 & NDCG@20\\
\toprule
CEDR~\cite{macavaney2019cedr} & - & .467 &  .538 \\
Birch~\cite{yilmaz2019cross} & .369 & .467 &  .532 \\
\midrule
BM25 & .253 & .363 & .424 \\
\ \ \  + T5-base & .314 & .425 & .510 \\
\ \ \  + T5-large & .296 & .416 & .499\\
\ \ \  + T5-3B & \textbf{.364} & \textbf{.506} & \textbf{.596}
\\
\midrule
BM25 + RM3 & .290 & .382 & .441 \\
\ \ \ + T5-base & .320 & .424 & .503 \\
\ \ \ + T5-large & .304	& .415 & .495 \\
\ \ \ + T5-3B & \textbf{.384} & \textbf{.510} & \textbf{.601} \\
\bottomrule
\end{tabular}
\end{center}
\vspace{2mm}
\caption{Results on Robust04. The T5 models are trained only on MS MARCO passage data and thus represent zero-shot transfer.}
\label{tab:robust04}
\end{table}

\begin{table}[t]
    \centering
    \begin{tabular}{l|c|c}
         &  2k samples & 20k samples \\
        \toprule
        BM25 + BERT-base & .127 $\pm${.058} & .201 $\pm${.012}\\
        BM25 + T5-base & .238 $\pm${.025} & .261 $\pm${.011}\\
        \bottomrule
    \end{tabular}
    \vspace{2mm}
    \caption{Comparisons between T5 and BERT trained with different numbers of training instances. Results report means and 95\% confidence intervals over five trials.}
    \label{tab:training_examples}
\end{table}

Results on Robust04 are shown in Table~\ref{tab:robust04}, where we apply our T5 reranker on top of retrieval results from BM25 and BM25+RM3 (see Section~\ref{section:training-and-inference}).
Figures in \textbf{bold} for T5-3B indicate that those results are significantly better ($p<0.05$) than T5-large, T5-base, and the corresponding baseline (BM25 or BM25+RM3), based on the Student's paired $t$-test with Bonferroni correction.
We compare our model with CEDR~\cite{macavaney2019cedr} and Birch~\cite{yilmaz2019cross}, two BERT-based state-of-the-art models.
Note that the CEDR results are from training on the Robust04 data (via cross-validation) and Birch uses Robust04 for tuning weighting parameters.
In contrast, we apply inference directly using our model trained on the MS MARCO passage data; Robust04 relevance judgments were only used as a test set, which makes our results zero-shot.
To our knowledge, our T5-3B model produces the highest known scores reported on Robust04.

As expected, effectiveness increases with larger models, but in all cases T5 is able to improve over both a bag-of-words as well as a query expansion baseline.
We explain the odd finding that T5-large performs worse than T5-base as follows:\ based on our training procedure, we simply ran a fixed number of iterations and then evaluated using the final checkpoint (see Section~\ref{section:training-and-inference}).
This has the advantage of not requiring validation data.
In the case of T5-large and T5-base, effectiveness does not monotonically increase in the out-of-domain dataset (Robust04), and thus the results reported in the table capture the somewhat arbitrary model state at the final checkpoint (where effectiveness may still be fluctuating within a rather large range).
For T5-3B, in contrast, effectiveness is far more stable across model checkpoints.
We leave for future work a more detailed examination of these model differences.
Although there are better ways of selecting model checkpoints that could lead to even higher scores, these techniques generally require cross-validation, which increases the danger of overfitting while abandoning the current (highly-desirable) zero-shot learning setup.

\subsection{Effect of Model Size and Training Data}
\label{sec:model_size_training_data}

Results from the MS MARCO passage ranking task (Table \ref{tab:msmarco}) represent a direct comparison between BERT and T5 models since the retrieval pipeline is otherwise the same.
For Robust04 (Table~\ref{tab:robust04}), we adopt a different architecture than CEDR and Birch, but effectiveness clearly improves as the size of the T5 model increases.
Therefore, while our T5-based approach achieves better results, it is entirely possible that the improvements are due to simply having a bigger model, as opposed to any intrinsic advantages over a classification-based approach.
Since we do not have pretrained T5 and BERT models of comparable sizes, it is difficult to conduct a fair empirical comparison.

Another interesting dimension of size is, of course, the amount of training data available.
In a data-poor regime with only a modest amount of training data, it appears that T5 can learn far more effectively than BERT.
To demonstrate this, we fine-tuned BERT-base and T5-base with either 1k (or 10k) positive and 1k (or 10k) negative instances sampled from the full MS MARCO passage dataset.
These two ``base'' models were selected due to their more modest computational demands for fine-tuning.
We trained them using a batch size of 32 for three epochs. For BERT, we used a learning rate $10^{-6}$ and no warm-up step. For T5, we used a learning rate of $10^{-3}$.
For each condition (2k or 20k samples in total), we repeated the experiment five times, drawing different samples each time.
The results are reported in Table~\ref{tab:training_examples}, with means and 95\% confidence intervals.

As expected, effectiveness significantly improves as we fine-tune the models with more data.
We see clearly that with the same amount of {\it limited} training data, T5 is significantly more effective than BM25.
In fact, with only 1k positive and 1k negative training instances, BERT performs worse than the BM25 baseline.
With 20k training instances in total, BERT is able to modestly improve upon BM25, but remains six points behind T5 fine-tuned on the same amount of data.
Interestingly, T5 is able to achieve roughly 45\% of the possible gain in effectiveness over the BM25 baseline with only 4\% of the training data.

\subsection{Target Word Probing Experiments}

Our experimental results immediately raise two questions:

\begin{enumerate}[leftmargin=*]

\item Why is our approach more data-efficient than BERT?
That is, why does T5 significantly outperform BERT when fine-tuned with few training examples?

\item How is our approach fundamentally different from classification, given that the softmax in our case reduces the model down to a binary decision?
That is, asking the model to decide between two output tokens seems no different from relevance classification.

\end{enumerate}

We believe these two issues are closely related.
Specifically addressing the second question:\
At a high level, both neural models are learning latent representations important to the task at hand (in this case, relevance classification), starting with a pretrained model, and then mapping these latent representations into task-specific decisions.
Thus, end-to-end task performance depends on a combination of the knowledge imparted via pretraining (already present at the start) and the knowledge gained via fine-tuning on task-specific data.
In the classification-based approach using BERT, the end-to-end model relies on a single fully-connected layer to map the latent representation (i.e., from the \texttt{[CLS]} token) into this binary decision.
While the approach can exploit pretrained knowledge when fine-tuning the latent representations, the final mapping (i.e., the fully-connected layer) needs to be, essentially, learned from scratch (since it is randomly initialized).

In contrast, T5 can exploit both pretrained knowledge and knowledge gleaned from fine-tuning in learning the proper task-specific latent representations as well as the mapping to relevance decisions.
Unlike the fully-connected layer in the classification-based approach, T5 can exploit the part of the network used for producing output.
Embedded in that neural machinery is latent knowledge about semantics, linguistic relations, and other features that are necessary to generate fluent text.
In other words, T5 has access to an additional source of knowledge that BERT does not.

This explanation, we believe, also answers the first question.
With plenty of training data, BERT has no trouble learning the final fully-connected layer (mapping latent representations to decisions), even from scratch (i.e., random initialization).
However, faced with few training examples, BERT still must learn the classification layer, but without any benefit from pretraining---and our experiments above (see Table~\ref{tab:training_examples}) show that it is unable to adequately do so.
In contrast, in a low-data regime, T5 can ``fall back'' on pretrained neural machinery used for generating fluent textual output.
In other words, our experiments suggest that the pretraining objective used in T5 can transfer well to {\it generating} relevance labels.

To turn our intuition into a testable hypothesis, we can vary the target words used as the prediction targets and manipulate their ``linguistic relatedness''---to deliberately ``disrupt'' linguistic knowledge that may be captured in the models.
As Puri and Catanzaro~\cite{puri2019zero} have shown, the choice of target words impacts effectiveness.
Recall that in our baseline, ``true'' indicates a relevant document and ``false'', a non-relevant document.
We tried the following contrastive variants:

\begin{itemize}[leftmargin=*]

\item {\bf ``Reverse''.} We swap the target words; that is, ``false'' indicates a relevant document and ``true'', a non-relevant document.
If the model is indeed exploiting latent knowledge about linguistic relations, then forcing the model to make opposite associations on the same polarity scale should lower effectiveness with respect to the baseline.

\item {\bf ``Antonyms''.} We map a relevant document to ``hot'' and a non-relevant document to ``cold''.
This preserves the use of adjectives at opposite ends of a polarity scale, but a scale that is completely unrelated to relevance.
If the model were exploiting latent knowledge, we would expect effectiveness to be lower than the baseline.

\item {\bf ``Related Words''.} We map a relevant document to ``apple'' and a non-relevant document to a related word ``orange''.
These words are semantically related, but do not present a polarity contrast as before.
We would expect effectiveness to be lower than the baseline.

\item {\bf ``Unrelated Words''.} We map a relevant document to ``hot'' and a non-relevant document to a completely unrelated word ``orange''.
Thus, we force the model to build an arbitrary semantic mapping.
We would expect effectiveness to be lower than the baseline and also lower than using related words.

\item {\bf ``Subwords''.} We map a relevant document to the subword ``\_ab'' and a non-relevant document to the subword ``\_de'' (note that we carefully select single tokens after tokenization by SentencePiece to avoid the need to combine multiple logits).
Here, we have removed all ``semantics'' from the input-to-output mapping.
We would expect effectiveness to be lower than the baseline and the above conditions.

\end{itemize}

Using these target word configurations, we conducted experiments on T5-base with either 1k (or 10k) positive and 1k (or 10k) negative instances sampled from the full MS MARCO passage dataset, same as in Section~\ref{sec:model_size_training_data}.
Once again, for each of the conditions, we repeated the experiment five times, drawing different samples every time.
For reference, we also fine-tuned with all available data.
We note that the effectiveness of T5-base is different from the one reported in Table~\ref{tab:msmarco} because we used slightly different hyperparameters which were more computationally efficient:\ here, we trained for 40k steps using a batch of size 256.
Experimental results are shown in Table~\ref{tab:analysis}, with means and 95\% confidence intervals.

\begin{table}[t]
    \centering
    \begin{tabular}{l|c|c|c|c|c}
        & \multicolumn{2}{c|}{Target Token} & \multicolumn{3}{c}{Training Size} \\
        Type & Positive & Negative & 2k & 20k & all \\
    \toprule
    Baseline & true & false & .238$\pm${.025} & .261$\pm${.011} & .355 \\
    \midrule
    Reverse & false & true & .222$\pm${.014} & .235$\pm${.021} & .340 \\
    Antonyms & hot & cold & .205$\pm${.032} & .216$\pm${.013} & .353 \\
    Related Words & apple & orange & .217$\pm${.024} & .234$\pm${.005} & .358 \\
    Unrelated Words & hot & orange & .202$\pm${.021} & .212$\pm${.001} & .351 \\
    Subwords & \_ab & \_de & .163$\pm${.027} & .151$\pm${.046} & .348 \\
    \bottomrule
    \end{tabular}
    \vspace{2mm}
    \caption{Results on the development set of the MS MARCO passage dataset comparing different target word manipulations.}
    \label{tab:analysis}
\end{table}

There does not appear to be an obvious pattern when fine-tuning with all available training data, although the largest observed difference is between ``baseline'' and ``reverse''.
This does appear consistent with our hypothesis that with sufficient training data, T5 is able to learn arbitrary mappings between document relevance and target words.
In the data-poor regime, the results are also consistent with our hypotheses.
With both 2k and 20k total samples, the baseline mapping achieves the highest effectiveness.
In the 2k condition, the confidence intervals computed from different samples mostly overlap (with the exception of subwords), so we do not have the benefit of greater certainty that comes with statistical significance.
On the 20k condition, our target word manipulations all significantly reduce effectiveness.
We note that the 95\% confidence intervals are smaller with more data, which illustrates the greater instability in effectiveness when training on smaller datasets (which is expected).

It is clear that the T5 model is taking advantage of latent semantic or linguistic knowledge in predicting relevance.
In both the 2k and 20k settings, the subwords condition performs worse than the BM25 baseline (and the 20k score is actually lower than the 2k score).
In this condition, T5 exhibits difficulty in achieving any predictive power at all.
There are at least two potential factors at play:\ we are removing all semantic associations, as the subwords are meaningless token fragments, and furthermore, we are forcing the model to produce tokens in an order (and context) that it has not encountered during pretraining.
We are unable to tease apart the effects currently, but either explanation is consistent with our intuitions.
For all other target word manipulations, we are at least able to beat the BM25 baseline.
Finally, our experiments are inconclusive regarding the importance of having a polarity scale in the low-data regime.
Quite clearly, reversing ``true'' and ``false'' has a large impact, but T5 is more effective learning targets that are semantically related but do not present a polarity contrast (``apple'' and ''orange'') than targets that encode an unrelated polarity contrast (``hot'' and ``cold'').

\section{Conclusion}

The main contribution of this paper is to introduce a novel generation-based approach to the document ranking task using pretrained sequence-to-sequence models.
Our models outperform a classification-based approach, especially in the data-poor regime with limited training data.
We attempt to explain these observations in terms of hypotheses about the knowledge that a model gains from pretraining vs.\ fine-tuning on task-specific data.
These hypotheses are operationalized into target word probing experiments, where we demonstrate that the model is indeed exploiting knowledge from its ability to generate fluent natural language text.
Exactly how remains an open research question and the focus of our ongoing work.

\section{Acknowledgments}

This research was supported in part by the Canada First Research Excellence Fund and the Natural Sciences and Engineering Research Council (NSERC) of Canada.
In addition, we would like to thank Google Cloud for credits to support this work.

\bibliographystyle{abbrv}
\bibliography{main}

\end{document}